\documentclass{aa}  

\usepackage{graphicx}
\usepackage{txfonts}
\newcommand{\teff}{$T_{\rm eff}$}

\newcommand{\msun}{${\rm M}_\odot$}

\newcommand{\rsun}{${\rm R}_\odot$}

\newcommand{\msunyr}{${\rm M}_\odot {\rm yr}^{-1}$}

\def\arcsec{$''$}
\def\mas{mas}
 
\def\kms{km~s$^{-1}$}

\def\teff{$T_{\rm eff}$}

\def\ha{${\rm H}\alpha$}

\def\ewha{$EW({\rm H}\alpha$)}
\def\brg{${\rm Br}\gamma$}

\def\macc{$\dot{M}_{\rm acc}$}

\def\rstar{${\rm R}_\star$}

\begin{document} 

   \title{Probing the magnetospheric accretion region of the young pre-transitional disk system DoAr 44 using VLTI/GRAVITY\thanks{Based on observations collected at the European Organisation for Astronomical Research in the Southern Hemisphere under ESO programme 0103.C-0097}}

%   \subtitle{I. A large-scale interferometric, spectropolarimetric, and multi-color photometric observing campaign\thanks{Based on observations obtained at ESO, CFHT, and LCOGT}}

 \titlerunning{DoAr 44 @ GRAVITY}
 \authorrunning{Bouvier et al.}

   \author{J. Bouvier
          \inst{1}
          \and K. Perraut\inst{1}
          \and J.-B. Le Bouquin\inst{1}
          \and G. Duvert\inst{1}
          \and C. Dougados\inst{1}
          \and W. Brandner\inst{2}
          \and M. Benisty\inst{1,3}
          \and J.-P. Berger\inst{1}
          \and E. Al\'ecian\inst{1}
          }
          
%          \fnmsep\thanks{Just to show the usage
%          of the elements in the author field}

   \institute{Univ. Grenoble Alpes, CNRS, IPAG, 38000 Grenoble, France\\
              \email{Jerome.Bouvier@univ-grenoble-alpes.fr}
         \and
         Max Planck Institute for Astronomy, K\"onigstuhl 17, 69117 Heidelberg, Germany
         \and
         Unidad Mixta Internacional Franco-Chilena de Astronom\'ia (CNRS, UMI 3386), Departamento de Astronom\'ia, Universidad de
Chile, Camino El Observatorio 1515, Las Condes, Santiago, Chile
%             \email{c.ptolemy@hipparch.uheaven.space}
%             \thanks{ESO}
             }

   \date{Received 29 January 2020; accepted 24 March 2020}

% \abstract{}{}{}{}{} 
% 5 {} token are mandatory
 
  \abstract
   {Young stellar objects are thought to accrete material from their circumstellar disks through their strong stellar magnetospheres. }
   {We aim to directly probe the magnetospheric accretion region on a scale of a few 0.01 au in a young stellar system using long-baseline optical interferometry.}
   {We observed the pre-transitional disk system DoAr 44 with VLTI/GRAVITY on two consecutive nights in the K-band. We computed interferometric visibilities and phases in the continuum and in the \brg\ line in order to constrain the extent and geometry of the emitting regions. }
   {We resolve the continuum emission of the inner dusty disk and measure a half-flux radius of 0.14 au. We derive the inclination and position angle of the inner disk, which provides direct evidence that the inner and outer disks are misaligned in this pre-transitional system. This may account for the shadows previously detected in the outer disk. We show that \brg\ emission arises from an even more compact region than the inner disk, with an upper limit of 0.047 au ($\sim$5~R$_\star$). Differential phase measurements between the \brg\ line and the continuum allow us to measure the astrometric displacement of the \brg\ line-emitting region relative to the continuum on a scale of a few tens of microarcsec, corresponding to a fraction of the stellar radius. }
   {Our results can be accounted for by a simple geometric model where the \brg\ line emission arises from a compact region interior to the inner disk edge, on a scale of a few stellar radii, fully consistent with the concept of magnetospheric accretion process in low-mass  young stellar systems.}


   \maketitle

%
%-------------------------------------------------------------------
    
\section{Introduction}

During the first few million years, the young low-mass stellar systems known as T Tauri stars are surrounded by a circumstellar disk from which they accrete material. In the inner disk region, at a distance of a few stellar radii above the stellar surface, the strong stellar magnetic field is able to disrupt the radial accretion flow, which is then forced to follow the magnetic field lines down to the stellar surface \citep{Bouvier07b, Hartmann16}. Magnetic funnel flows thus develop,  connecting the inner disk edge to the stellar surface on a scale of a few 0.01 au. At the distance of the nearest star forming regions ($\geq$ 100~pc), the angular size of the star-disk interaction region is therefore of the order of a milli-arcsec (mas) or less, a scale barely reachable with even the largest telescopes. 
This is the reason why the magnetospheric accretion process in T Tauri stars has been investigated mostly indirectly so far, for example, by monitoring the variability of the inner system as it rotates \citep[e.g.,][]{Bouvier07a, Alencar12, Alencar18,  Donati19}. However, interferometric facilities offer the promise to reach these scales \citep{Eisner09, Eisner10, Eisner14}, and the spectacular improvement in sensitivity and stability of newly available interferometric near-infrared instruments now allows us to directly probe the accretion flows within the magnetosphere of young stellar systems. We report here such an attempt using GRAVITY at the VLTI  \citep{Grav17}.

%{\bf KP/JB: gain in sensitivity in near-ir interferometry in the last years allows us to explore...}

We performed a large-scale, multi-instrument campaign in June 2019, monitoring the young pre-transitional disk system DoAr 44 (aka V2062 Oph, Haro 1-16, ROXs 44, HBC 268). This young stellar object is a 1.2~\msun\ classical T Tauri star (cTTS) located in the Rho Ophiuchus dark cloud at a distance of 146$\pm$1~pc \citep{Gaia16, Gaia18}. A moderately bright source of spectral type K3, it exhibits strong \ha\ emission   \citep[\ewha$\simeq$50\AA,][]{Bouvier92} and accretes at a substantial rate from its circumstellar disk \citep[\macc =  6.0-9.3 10$^{-9}$ \msunyr,][]{Espaillat10, Manara14}. From the measurement of Zeeman broadening of near-infrared FeI lines, \cite{Lavail17} derived a mean surface magnetic field strength amounting to 1.8$\pm$0.4~kG, and possibly up to 3.6~kG. The moderate mass accretion rate and strong magnetic field makes DoAr~44 a prime target to investigate the magnetospheric accretion process in this system.

The overall results of the multi-instrument campaign are reported in an accompanying paper (Bouvier et al., in prep.; hereafter Paper I). Here, we present and analyze the unique datasets obtained from long baseline optical interferometry, using VLTI/GRAVITY. Section~2 describes the observations. Section 3 presents the analysis of the interferometric visibilities and phases both in the K-band continuum and across the \brg\ emission-line profile. Section~4 provides an interpretation of the results in terms of a simple geometrical model for the star-disk interaction region in this system, based on the interferometric observables.

%
%--------------------------------------------------------------------
%--------------------------------------------------------------------   
%--------------------------------------------------------------------
\section{Observations}

%-------------------------------------- Two column figure (place early!)

        \begin{table*}[t]
\caption{Journal of the VLTI/GRAVITY observations. $N$ denotes the number of five-minute-long sequences that have been recorded on the target.}
\label{gravity_journal}      % is used to refer this table in the text
\centering
\vspace{0.1cm}
\begin{tabular}{l l l l l l l l}
\hline \hline
HJD & Date & Time & Configuration & N & Seeing & $\tau_0$ & Calibrators \\
& &(UT) & & & (\arcsec) & (ms)\\
\hline
 2458656.66339 & 2019-06-22 & 02:04 - 05:46 & UT1-UT2-UT3-UT4 & 26 & 0.41 - 0.88 & 6.4 - 15.8 & HD~147578, HD~149562\\
2458657.59764  & 2019-06-23 & 01:48 - 02:50 & UT1-UT2-UT3-UT4 & 10 & 0.84 - 1.16 & 6.8 - 9.7 & HD~146706, HD~146235\\
\hline
\end{tabular}
\end{table*}

   \begin{figure*}[t]
   \centering
    \includegraphics[width=\hsize]{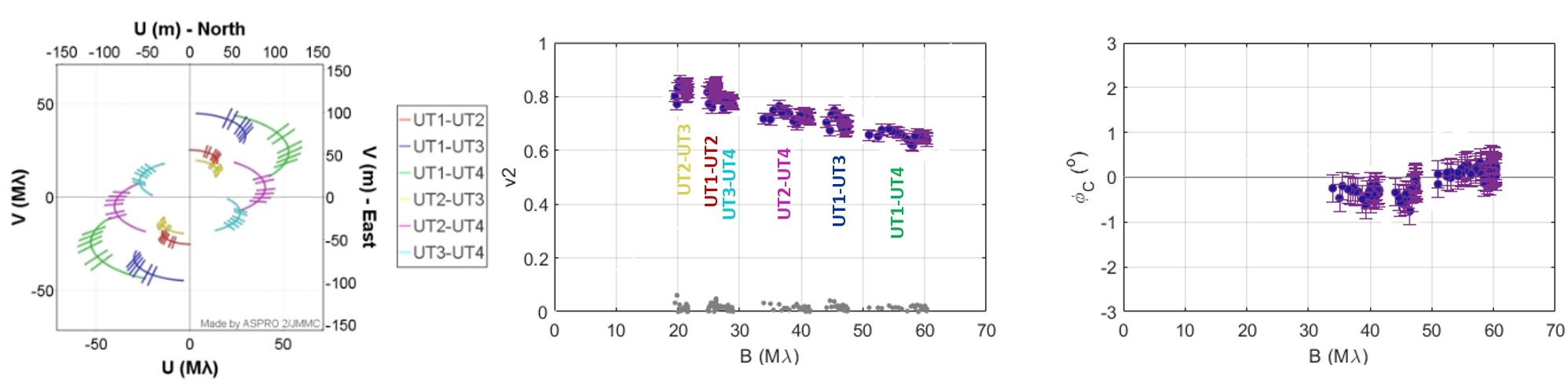}
   \caption{GRAVITY observations of DoAr~44 on June, 22. {\it Left:} (u, v) spatial frequency plane corresponding to the six projected baselines during the 3.7 hours of observations. {\it Middle:} continuum squared visibilities as a function of baseline, recorded in the central spectral channel of the fringe tracker ($\lambda$~=~2.15~$\mu$m). The gray circles at the bottom of the panel correspond to the residuals after the best fit model has been subtracted (see text). {\it Right:} closure phases measured in the central spectral channel as a function of baseline.  The error bars amount to 2\% on the squared visibilities and 0.3$\degr$ on the closure phases (see text for details).} %
              \label{visibilities}%
    \end{figure*}
    
We observed DoAr~44 on June 22 and 23, 2019 in the K-band with the GRAVITY instrument \citep{Grav17}, combining the four unit telescopes of the ESO/VLTI (ESO run 60.A-9256). At 2.2~$\mu$m, with a maximum baseline of 130~m, we thus reach an angular resolution of $\lambda$/2$B_{\rm max}=1.7$~mas, corresponding to 0.25 au at 146 pc. 
After the MACAO adaptive optics correction, the flux of the source is split and feeds the fringe tracker and science instrument simultaneously.
%DoAr~44 simultaneously fed the fringe tracker, the MACAO Adaptive Optics, and the science instrument. 
We collected exposures with coherent integrations of 0.85\,ms on the fringe tracker to freeze the atmospheric effects \citep{Lacour19}. Once the fringes were stabilized, we recorded successive sequences of 5 minutes, namely 10 exposures of 30\,s each, on the object with the science instrument in its high spectral resolution mode (R$\sim$4,000). The datasets thus consist of long sequences of integration on DoAr~44, interleaved with sky exposures and interferometric calibrator observations. The calibrators were chosen to be unresolved at the longest baselines, located close to the target on the sky, and having a similar magnitude in the K band. Several calibrators were observed in order to derive the instrumental transfer function accurately. Weather conditions were excellent during the first night and we recorded 26 files of five minutes each over a period of 3.7 hours. At the start of the second night, we recorded only one hour of exploitable data, as the domes had to be closed to high winds thereafter. The journal of observations is given in Table~\ref{gravity_journal}.

%--------------------------------------------------------------------
%--------------------------------------------------------------------
%--------------------------------------------------------------------

\section{Results}

We report in this section the data reduction processes that allowed us to measure interferometric visibilities and phases along the six baselines sampled by the four ESO/VLTI unit telescopes. From these measurements, we derive the main properties of the emitting regions, both in the continuum and across the \brg\ line profile.

\subsection{Continuum visibilities}
\label{sec:continuum}

%{\bf KP/JPB: Describe ellipsoid model; mention that we also tried a ring and yield the same results}.

We used the standard GRAVITY pipeline \citep{Grav17, Lapeyrere14} to reduce the fringe tracker data. Complex visibilities were computed over five spectral channels and were used to constrain the extent of the K-band continuum emitting region. 
Since the error bars provided by the pipeline can be underestimated and/or do not include the residual calibration effects, we adopted conservative error bars: for the squared visibilities, we computed the rms over exposures probing the same spatial frequencies, which yields error bars amounting to 2\% in all spectral channels; for the closure phases, we computed the rms over the 36 exposures recorded during the two nights and derived an rms of 0.8$\degr$ when considering all the spectral channels, and of 0.3$\degr$ when considering the central spectral channel only. The squared visibilities can be fit as a function of the baseline with geometrical models, 
%and their wavelength dependence provides the spectral slope of the circumstellar environment. 
 while the closure phase measures the degree of asymmetry of the emitting region \citep[e.g.,][]{Monnier07}. Figure~\ref{visibilities} displays results from the dataset obtained on June 22, 2019. The K-band continuum-emitting region of DoAr~44 appears to be partially resolved ($V^2$ < 1) and  centro-symmetric as suggested by a closure phase close to 0$\degr$. The results obtained from the observations performed on June 23, 2019 are consistent with the results of the first night but do not significantly add to the signal to noise of the visibilities or phases due to their limited duration. 

To model the visibilities, we followed the approach \citet{Lazareff17} and \citet{Gravity19} applied to the analysis of interferometric surveys of Herbig stars performed with PIONIER and GRAVITY, respectively. For the continuum K-band emitting region, we fit the visibility measurements as a function of baseline with a three component model: a point-like central star (since we are not able to resolve the stellar photosphere), a circumstellar disk that is modeled as a 2D elliptical Gaussian, and an extended component. The latter contribution, also called halo in \citet{Lazareff17}, is assumed to be fully resolved, leading to visibilities smaller than 1 at very short baselines. This extended component contribution is sometimes interpreted as scattered light \citep{Pinte2008}. Indeed, it has been found to improve the parametric modeling of young stellar objects for which the scattered light flux is significant, mainly transitional disks \citep{Lazareff17} and T Tauri stars \citep{Anthonioz15}. 
The free parameters are the relative contributions to the total flux of the star $f_s$, the circumstellar disk $f_c$, and the extended component (if any) $f_h$, with the sum of these contributions being equal to unity, the half-flux semi-major axis, the inclination, and the position angle of the circumstellar disk. The total complex visibility at the spatial frequency (u, v) and wavelength $\lambda$ is given by:
\begin{equation}
    V(u, v, \lambda) = \frac{f_s(\lambda) + f_c (\lambda) V_c(u, v)}{f_s(\lambda) + f_c (\lambda) + f_h (\lambda)},
\end{equation}
where $V_c$ denotes the visibility of the circumstellar disk, and the wavelength dependence of the flux contributions is explained in \citet{Lazareff17}\footnote{The chromatic exponents from \cite{Lazareff17} model amount to $k_s$=1.1 and $k_c$=-0.2 in our best fit model.}.

%the spectral index of the circumstellar environment, 
%and the weighting scheme for the radial brightness distribution [LAISSE-T-ON CE PARAMETRE ?]. The latter is allowed to vary from a purely Gaussian to a purely Lorenzian distribution.} %The ring model has the same set of parameters with the additional width of the ring. 

Since the disk is only partially resolved by our observations, its flux contribution $f_c$ and half-flux radius are partly degenerated \citep{Lazareff17}. We used the K-band excess ratio $F_{exc}/F_\star$= 0.8~$\pm$~0.2 measured by \citet{Espaillat10} from the spectral energy distribution of the system to derive a starting value for $f_c$ but kept it as a free parameter for the visibility fit. The flux contributions we derive for the different components indicate that, at the time of our observations, the star dominated the continuum emission in the K-band, with a disk fractional flux $f_c$ ranging from 20 to 34\%. The best fit model yields  $F_{exc}/F_\star$=0.49$\pm$0.12, where the difference from \citet{Espaillat10} estimate may reflect intrinsic variability of the system over a timescale of years.
The visibilities corresponding to the best fit model do not reach unity at zero baseline, which is interpreted as the presence of an additional, fully resolved extended component that contributes about 6-7\% to the total flux of the system in the K-band. The results of the best-fit model are listed in Table~\ref{gravityfit}.  The 1-$\sigma$ errors are provided for all the parameters, and take into account the partial degeneracy between the disk fractional flux and the half-flux radius. 
%. For the latter, we provide the mean and the standard deviation over the range of flux variation. 
%Regardless of the flux starting value, 

The model converges towards a 2D elliptical Gaussian whose minor-to-major-axis ratio is equal to 0.83, corresponding to an inclination of 34$^\circ$, at a position angle of 140$^\circ$ measured east from north. The mean half-flux radius of 0.96 mas corresponds to 0.14~au ($\sim$15~\rstar)\footnote{The stellar radius, \rstar=2.0$\pm$0.15 \rsun,  was obtained in Paper I from the luminosity and \teff\ we derived for the star.} at a distance of 146~pc. %When the disk fractional flux varies from 20 to 34\%, the half-flux radius varies between 1.0 and 0.85 mas. 
We checked that the data obtained on the second night, in spite of their lower quality, are in agreement with these parameters. We also checked that a ring model converges towards the same values of parameters, which indicates that we are not able to resolve an inner cavity at the scale of our angular resolution.

%The parameters of the best fit model are listed in Table~\ref{gravityfit}. The ellipsoid model converges towards an almost symmetric 2D Gaussian brightness distribution with a half-flux radius of 0.8~\mas, corresponding to 0.12 au (12.6~\rstar)\footnote{The stellar radius, \rstar=2.0$\pm$0.15 \rsun,  was obtained in Paper I from the luminosity and \teff\ derived for the star.} at a distance of 146~pc. From the  of the best ellipsoid model, we further derive an inclination of 34$^\circ$ and a position angle of about 140$^\circ$ for the major axis. We also checked that the data obtained on the second night, in spite of their worse quality, are in agreement with these parameters. %

%This is in qualitative agreement with the K-band excess ratio $F_{exc}/F_\star$= 0.8~$\pm$~0.2 measured by Espaillat+2010 from the spectral energy distribution of the system. The half-flux radius derived above is sensitive to the disk to total flux ratio at 2.2~$\mu$m. With a disk fractional flux varying from 25 to 31\%, the half-flux radius changes from 1 to 0.85 mas. For each model, the 1-$\sigma$ error bar is of order of 6-7\% on the fractional disk flux, which translates to 0.1 mas error on the half-flux radius. Finally, the visibilities corresponding to the best fit model do not reach unity at zero baseline, which is interpreted as the presence of an additional, fully resolved extended component that contributes about 6-7\% to the total flux of the system in the K-band. 

\begin{table}
\caption{Best-fit parameters of the K-band continuum VLTI/GRAVITY data of DoAr~44 obtained on June 22, 2019. See text for a detailed description of the model. }
\label{gravityfit}
\begin{center}
\vspace{0.1cm}
\begin{tabular}{l l}
\hline \hline
Parameters & Ellipsoid \\
\hline
$f_c$ & 0.27~$\pm$~0.07 \\
$f_h$ & 0.06$\pm$~0.01 \\
$f_s$  & 0.67$\pm$~0.07 \\
$\cos i$ & 0.83$\pm$~0.02 \\
$PA$ [$^\circ$] & 140~$\pm$~3 \\
%Radial brightness distribution$^{(a)}$ & 0.16~$\pm$~0.13 \\
Half-flux radius [\mas] & 0.96~$\pm$~0.18 \\
Half-flux radius [au] & 0.14~$\pm$~0.026 \\
Half-flux radius [\rstar] & 15.0 $\pm$ 2.8\\ 
%Ring width  $w^{(b)}$  & -- & 0.98~$\pm$~0.02 \\
%$T_{\rm dust}$ & tbc & tbc \\
$\chi_r^2$ & 1.06 \\
\hline
\end{tabular}
\end{center}
\end{table}

%{\bf Continuum modeling: half flux radius = 0.8mas = 0.12 au !!! BrG visibilities higher, i.e., even more compact !!!}

\subsection{Br$\gamma$ line interferometric observables}
\label{sec:brG_fit}

The spectrally dispersed observations from the science instrument (SC) were reduced with the latest version of the instrument pipeline. It includes recent fixes to lower the imprint of the internal source in the calibration frames, and a better filtering of the outliers created by cosmic rays. As a sanity check, we reduced the SC data with four independent sets of calibration files, recorded on June 18, 20, 22, and 27. We obtained consistent results for all of them. 

The calibrator observations were somewhat noisier than the science data, due to the limited number of calibration sequences. We therefore did not perform an absolute calibration of the spectrally dispersed observations, as this would have degraded the signal-to-noise ratio of the science data. Instead, we used differential measures that can be reliably computed by renormalizing the spectrally dispersed data, forcing the continuum around the Br$\gamma$ line to match the prediction of the best-fit model of the continuum emission derived from the calibrated FT data (see Sec.~\ref{sec:continuum}). We averaged all observations for each of the two nights in order to increase the signal-to-noise ratio. The subsequent $uv$-plane smearing has no significant impact on the results because the relevant scales are marginally resolved. The spectrally dispersed interferometric observables are shown in Figure~\ref{fig:BrG}. The first night has typical standard deviations of 0.004 in visibility amplitude, 0.25$\degr$ in phase for each baseline, and 0.45$\degr$ in phase closures, while the second is of lower quality, with 1-$\sigma$ uncertainties of 0.01 in amplitude, 0.5$\degr$ in phase, and 0.8$\degr$ in phase closures.
%{\bf JB: Explain how this is computed.}

\begin{figure}
  \centering
   \includegraphics[width=\columnwidth]{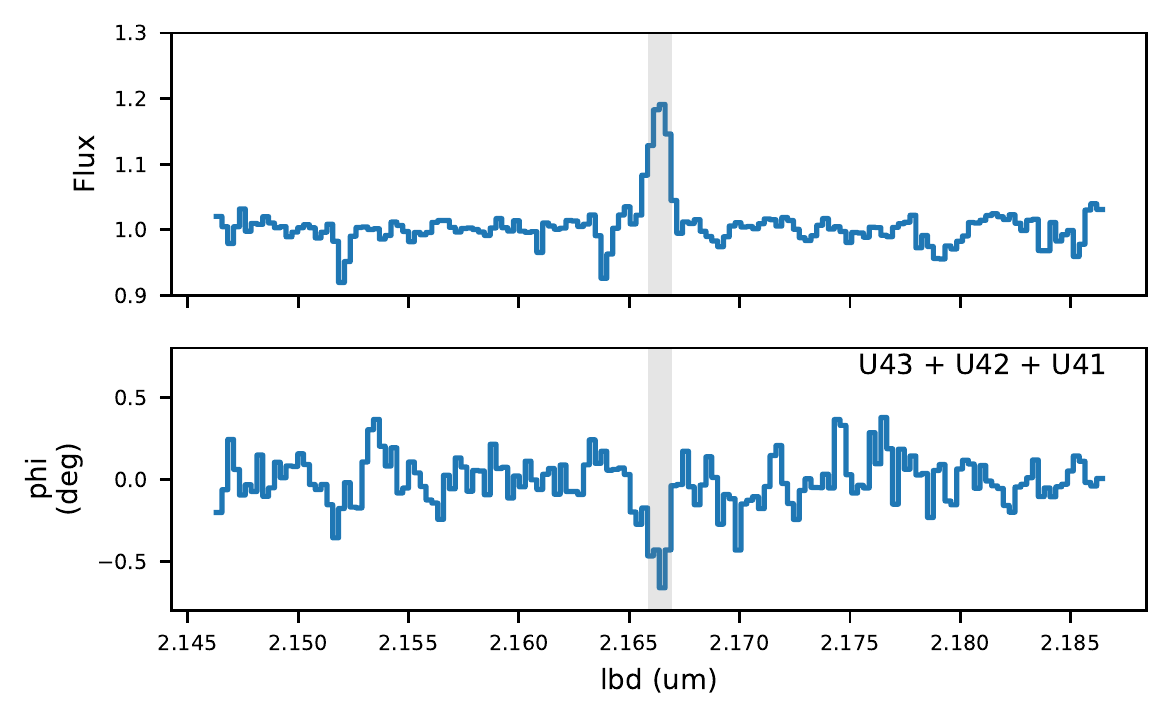}
  \caption{{\it Top panel:} Spectrum around the Br$\gamma$ line for the night 2019-06-22. The gray area shows the extent of the \brg\ line profile.  {\it Bottom panel}: Mean of the differential phase of the three  baselines UT4-UT3, UT4-UT2, and UT4-UT1 for the night 2019-06-22. The standard deviation of the summed signal is 0.16$^\circ$. The averaged phase signal is detected across the line profile at the 4~$\sigma$ level.}
  \label{fig:sumphi}%
\end{figure}

The \brg\ line appears in emission in the spectrum. Its measured FWHM is 1.25\,nm, therefore slightly enlarged compared to the 0.40\,nm spectral resolution of GRAVITY. This broadening corresponds to a velocity dispersion of 164~\kms. An increase of visibility amplitude is seen across the line profile for all baselines. A small but consistent signal is also seen in the differential phase across the line profile along three baselines over the two nights (see Fig.~\ref{fig:BrG}). It reaches a level of 2.2, 2.5, and 3.3~$\sigma$ along baselines UT4-3/2/1, respectively, for the first night. When averaging the three baselines, the signature appears at the 4~$\sigma$ level across the line profile, as shown in Figure~\ref{fig:sumphi}. There is no significant detection of a  phase closure signal across the \brg\ line at a 1$\sigma$ level of 0.31, 0.43, 0.31, 0.25~deg for the first night. The signatures of all interferometric observables across the line profile are adequately modeled by a Gaussian, with no evidence for a more complicated spectral profile (e.g., double peaked or S-shaped), and the behavior of interferometric observable is the same for the two nights. 

The simplest geometrical model able to account for such signatures is a compact, mostly unresolved \brg\ emission region. It is hereafter described by a Gaussian, without spatially resolved kinematics, which is slightly offset with respect to the barycenter of the continuum emission. In addition to the \brg\ line position, broadening, and flux, the free parameters of the model are the angular half-flux radius of the \brg\ emitting region, its astrometric offset relative to the continuum emission, and the position angle of this offset. While adjusting the spectrally dispersed data with the model, we kept the parameters of the continuum fixed since they are constrained by the FT data onto which the continuum has been rescaled. The best-fit model is overlaid to the data in Figure~\ref{fig:BrG}, and the corresponding best-fit parameters are listed in Table~\ref{tab:BrG}.

%The Br$\gamma$ line appears in emission in the spectrum. An increase of visibility amplitude is seen across the line profile for all baselines. A small but consistent signal in phase is also seen across the line profile along 3 baselines (UT4-3/2/1) over the two nights. The behavior of interferometric observables is the same for the two nights. The simplest geometrical model able to account for such signatures indicates a compact Br$\gamma$ line emitting region, slightly offset with respect to the barycenter of the continuum emission. The free parameters of the model are: the Br$\gamma$ line flux, the angular half-flux radius of the Br$\gamma$ emitting region, its offset relative to the continuum emission, and the position angle of the astrometric offset. 

\begin{table}[t]
  \caption{Best-fit parameters of the Br$\gamma$ VLTI/GRAVITY data of DoAr~44. The errors are 1-$\sigma$ uncertainties.}
  \label{tab:BrG}
  \centering
  \vspace{0.1cm}
  \begin{tabular}{l l l}
    \hline \hline
    Parameters & 2019-06-22 & 2019-06-23 \\
    F$_{\mathrm{Br}\gamma}$\,/\,F$_{cont}$ & $0.19\pm0.02$ & $0.17\pm0.02$ \\
    Half-flux radius [mas] & $<0.32$ & $<0.25$\\ 
    Half-flux radius [au] & $<0.047$ & $<0.037$\\ 
    Half-flux radius [\rstar] & $<5.0$ & $<3.9$\\ 
    Offset [$\mu$as] & $52\pm15$ & $67\pm55$ \\
$PA$ [$^\circ$] & $163\pm15$ & $158\pm50$ \\
    \hline
  \end{tabular}
\end{table}

The Br$\gamma$ emitting region is not spatially resolved. The upper limit of 0.32 mas we derive on its size corresponds to 0.047\,au at a distance of 146\,pc, which translates into 5~\rstar, for a stellar radius of 2~\rsun, as derived in Paper I.  The Br$\gamma$ line emitting region is thus significantly more compact than the inner circumstellar disk whose half-flux radius reaches 15~\rstar. The astrometric offset measures the distance between the photocenter of the \brg\ emitting region and that of the continuum. It is listed in Table~\ref{tab:BrG} and is expressed in microarcseconds, illustrating the exquisite precision of differential near-infrared long baseline interferometry. For the first night, the offset value of $52\,\mu$as corresponds to 0.0076~au, meaning  0.82\,\rstar, at a position angle of about 160$\degr$. The poorer signal-to-noise ratio gathered during the second night unfortunately prevents us from detecting any temporal evolution of either the amplitude or the position angle of the astrometric offset.

% {\bf State results: [KP, JB, GD]:}
% Upper limit on the size of the Br$\gamma$  emitting region = 0.4 mas (d=146pc, 0.058 au = 12.6 Rsun = 6.3 Rstar)

% Modelling the differential phase one gets:
% MJD = 58656.087  sep = 0.053 mas  (0.0077 au = 1.67 Rsun = 0.84 Rstar);  PA = 162.8 deg; Phirot=0.77
%  MJD = 58657.075  sep = 0.061 mas  (0.0089 au = 1.92Rsun = 0.96 Rstar);PA = 159.2 deg; Phirot=0.08
% Compare to S CrA Gravity results.
 
\section {Discussion}
DoAr~44 belongs to the class of relatively rare young stellar systems with pre-transitional circumstellar disks. 
 On large scales, high angular resolution 880~$\mu$m continuum maps obtained with SMA by \cite{Andrews09, Andrews11} reveal the existence of a dust-depleted cavity extending up to a distance of 30~au from the central star with the cold dust emission arising from an outer ring about 25 au-wide. ALMA observations in the optically thin $^{13}$CO and C$^{18}$O lines reveal a ring structure similar to that seen in the dust continuum emission albeit somewhat wider, namely a rotating gaseous ring extending from about 25 to 60~au from the central star, seen at an inclination of $\approx$20\degr \citep{vanderMarel16}. In addition, adaptive optics H-band polarized images reveal the presence of two symmetric shadows in the azimuthal brightness profile of the outer ring seen in scattered light \citep{Casassus18}. A radiative transfer model suggests that these shadows are cast by an inner disk warp, with a relative inclination amounting to 30$\pm$5\degr\ between the inner disk and the outer ring. On small scales, \cite{Espaillat10} ascribed the near-infrared excess observed in the spectral energy distribution of the source to a hotter dust component, located at the dust sublimation radius of an optically thick inner disk. Using a radiative transfer model to describe the inner disk wall, they deduce a dust sublimation radius of 0.25~au at a temperature of 1200~K, and a size less than 0.4~au for the inner disk. \cite{Salyk09, Salyk11} investigated the inner disk gaseous component by modeling the rovibrational transitions of CO lines seen in emission in the near-IR spectrum of the source. They derived a gaseous disk inner radius of 0.3$\pm$0.1~au, which is consistent with that of the dusty component responsible for the near-IR excess. Thus, pre-transitional disk systems consist of a compact inner disk close to the star surrounded by a wide gas and dust-depleted cavity, and an outer ring or disk component, with a possible misalignment between the inner and outer scales of the circumstellar material.  

Near-IR spectro-interferometry and spectro-astrometry are powerful tools to explore the innermost regions of young stellar objects surrounded by circumstellar disks \citep[e.g.][]{Malbet07, Kraus08, Goto12, Eisner14, Mendigutia15, Gravity17, Gravity19, Gravity20}. Indeed, our long baseline optical interferometric observations allow us to directly constrain the properties of the inner system. The inner dusty disk of DoAr 44 is resolved in our GRAVITY data with a half-flux radius of 0.14 au, corresponding to $\sim$15~\rstar. This is nearly a factor of two more compact than the location of the 1200~K inner dust wall inferred by \cite{Espaillat10} to account for the near infrared excess flux of DoAr~44's spectral energy distribution. Given uncertainties attached to the latter estimate, which depends upon the assumed inner wall geometry, the dust properties, and the stellar and accretion properties, the discrepancy might either not be significant or could indicate long-term variations in the extent of the star-disk interaction region. From the fit of the continuum visibilities, we further derive an inclination of 34$\degr$$\pm$2$\degr$ and a PA of 140$\degr$$\pm$3$\degr$ for the inner disk. These values are remarkably consistent with the inclined disk model inferred by \cite{Casassus18}, which requires $i\simeq$29.7$\degr$ and PA$\simeq$134$\degr$ for the inner disk to account for the location of the shadows seen in the outer dusty ring. Hence, the GRAVITY results presented here seem to provide direct support to the interpretation of shadows seen in the outer circumstellar ring as being due to a misalignment between the inner and outer disks in this pre-transitional disk system. 

%The rotational period and projected rotational velocity of the DoAr~44 system we derive in Paper I yield an inclination of  30$\pm$5$\degr$ for the stellar rotational axis onto the line of sight. We thus conclude that the inner disk and the central star have similar inclinations, while on a scale of several tens of au the outer disk is misaligned relative to the inner system. 

The larger visibilities we measure across the \brg\ line profile compared to those computed for the nearby continuum readily indicate that the \brg\ emitting region is more compact than the inner dusty disk. The model fit of the interferometric visibilities yields an upper limit of 0.047 au, meaning 5~\rstar, to the \brg\ line-emitting region. This is typically the size of the magnetospheric accretion region in young stellar systems \citep[e.g.,][]{Bouvier07b} and quite similar to the corotation radius of 0.043 au (4.63~\rstar) we derived for the system in Paper I (P$_{rot}$=2.96~d).
%(P=2.96 d, \mstar=1.2\msun, \rstar=2.0\rsun). 
Moreover, the differential phase signal indicates a small but significant astrometric offset between the photocenter of the line-emitting region relative to that of the continuum emission. 
%of the system, assumed to correspond to the center of the stellar photosphere\footnote{This is assuming a similar contribution of the star and the disk at 2.2$\mu$m (see Table~\ref{gravityfit}) and an axisymmetric inner dusty disk, as supported by the lack of a closure phase signal in the interferometric measurements}. 
The offset amounts to 0.0076 and 0.0098 au, corresponding to 0.82 and 1.06~\rstar, on the first and second nights, respectively, and lies at a position angle of about 160$\degr$. Within significant uncertainties (cf. Table~\ref{tab:BrG}), the amplitude of the astrometric offset and its position angle may have remained constant over this timescale. 

Do these results fit the magnetospheric accretion paradigm for low-mass young stellar objects? The \brg\ line flux in the spectra of young stars has been shown to diagnose and reliably measure the mass-accretion rate onto the star \citep[e.g.,][]{Najita96, Muzerolle98a}. Radiative transfer modeling of line emission arising from magnetospheric funnel flows in an aligned dipolar geometry predicts relatively symmetric \brg\ line profiles \citep{Muzerolle98b}. At low inclination, as is relevant for DoAr 44 ($i\simeq30\degr$, Paper I), the profile does not exhibit conspicuous inverse P Cygni features, and is nearly centered at the star's velocity \citep{Kurosawa11}. All these properties are consistent with the Gaussian-like, centered \brg\ line profile we report here at moderate spectral resolution (see Fig.~\ref{fig:sumphi}). 

The near-IR hydrogen line emission is expected to arise from the bulk of the magnetosphere, extending from the disk truncation radius down to the stellar surface \citep{Kurosawa08}. Assuming the truncation radius is close to the corotation radius, as is often the case for T Tauri stars \citep[e.g.,][]{Alencar12, Alencar18, Donati13, Donati20}, we expect \brg\ emission to arise over a large volume around the star, extending up to about 4.6~\rstar. This is consistent with the upper limit we derive on the half-flux radius of the \brg\ line-emitting region  from interferometric visibilities (see Table~\ref{tab:BrG}).

Even in the case of a slightly misaligned dipole\footnote{We derive a 20$\degr$ magnetic obliquity for DoAr 44 in Paper I from Zeeman Doppler Imaging.}, line emission from magnetospheric funnel flows remains relatively symmetric at low inclination \citep[see, e.g., model A in][]{Kurosawa08}.
%thus accounting for the lack of significant phase closure in the \brg\ line. 
The astrometric offset we measure between the photocenter of the  \brg\ emission region and the center of the stellar disk amounts to a fraction of the stellar radius (0.6-1.1~\rstar, see Appendix B). While a full MHD-radiative transfer model would be required to clarify the origin of such a small offset, we conjecture that it could result from an azimuthal modulation of the funnel flow emission, as expected from accretion, onto a slightly misaligned dipole or a more complex large-scale magnetic topology indeed \citep[e.g.,][]{Kurosawa13}.  

Overall, we therefore believe that our results are consistent with the assumption that we might be directly probing the \brg\ emitting region arising from funnel flows within the stellar magnetosphere of the young DoAr 44 system.

Alternatively, bipolar stellar winds or jet outflows may also be expected to contribute near-IR hydrogen line emission in young stellar systems \citep[e.g.,][]{Kurosawa12}, in a direction perpendicular to the disk midplane. However, we find here that the PA of the astrometric offset of the \brg\ emission is nearly along the disk's major axis, which does not support an interpretation in terms of polar outflows. The base of disk winds is another potential source of \brg\ emission but it would have to originate here within 5~\rstar\ from the central object. Moreover, rotating disk winds and bipolar outflows alike, yield an S-shape signature in the spectrally dispersed differential interferometric phases across the line profile \citep[e.g.,][]{Weigelt11, Kreplin18}, which is not seen here (see Fig.~\ref{fig:sumphi}).  

Finally, more exotic interpretations cannot be ruled out. For instance, part of the \brg\ emission could arise from an accreting planet orbiting close to the star. Hydrogen emission has been directly detected in the circumstellar disks of young systems at the location of wide-orbit planets  \citep[e.g.,][]{Haffert19}. Compact close-in planetary systems, such as those revealed by Kepler with orbital periods of a few days \citep[e.g.,][]{Winn18}, could conceivably yield these kinds of signatures when still embedded in the inner disk. Additional evidence, such as periodic radial velocity variations, would however be required to support this hypothesis. We do detect a periodic modulation of the star's radial velocity in Paper I, with an amplitude of about 0.6~\kms, but the modulation occurs at the stellar rotation period, which points to stellar activity.

\section {Conclusion}

Long baseline near-infrared interferometry provides a unique opportunity to get direct constraints on the inner scales of the environment of young stellar objects, down to a few 0.01 au. Its application to the young pre-transitional disk system DoAr 44 has allowed us to derive the properties of the inner disk and to put strong constraints on the origin and size of the \brg\ line emitting region. 

We resolve a compact inner disk on a scale of 0.14 au around the central star, and derive similar inclinations onto the line of sight for the inner disk and the central star, which is different from that of the outer ring seen at much larger scales. This result provides strong support to the interpretation of dark shadows seen in outer rings of pre-transitional disk systems as resulting from inner-outer disk misalignment. 

The interferometric observables across the \brg\ line profile reveal a compact line emitting region, of the order of 0.04 au or less. Thus, both the maximum size of the \brg\ emitting region, of the order of 5\rstar\ or less, and its offset from the central star, which amounts to a fraction of the stellar radius, suggest that the \brg\ emission arises in the funnel flows of the magnetospheric accretion region, located between the inner disk edge at 0.12 au and the stellar surface. Indeed, the maximum size we derive for the \brg\ line-emitting region is similar to the Keplerian disk's corotation radius, which usually coincides with the radius at which the inner disk is truncated as it encounters the stellar magnetosphere.  

Finally, we wish to emphasize that as sharp as it is, the interferometric view of the inner regions of young accreting systems is best exploited when combined with complementary approaches, such as spectropolarimetry, high-resolution spectroscopic monitoring, and multicolor photometry. The interferometric results presented here are further discussed in the context of the full observing campaign on DoAr 44 reported in Paper I. It is only through the simultaneous use of these powerful techniques that we may ultimately hope to fully decipher the physical processes taking place on a scale of a few 0.1 au or less around young stars, indeed the birth site of the wealth of compact inner planetary systems the Kepler satellite has unveiled over the last years. 

\begin{acknowledgements} 
We would like to thank all the individuals who have contributed to build such a powerful instrument as GRAVITY. We thank the referees for their reports that improved the accuracy and clarity of the manuscript. This project has received funding from the European Research Council (ERC) under the European Union's Horizon 2020 research and innovation programme (grant agreement No 742095; {\it SPIDI}: Star-Planets-Inner Disk-Interactions, http://www.spidi-eu.org). This work has made use of data from the European Space Agency (ESA) mission
{\it Gaia} (\url{https://www.cosmos.esa.int/gaia}), processed by the {\it Gaia}
Data Processing and Analysis Consortium (DPAC,
\url{https://www.cosmos.esa.int/web/gaia/dpac/consortium}). Funding for the DPAC
has been provided by national institutions, in particular the institutions
participating in the {\it Gaia} Multilateral Agreement.
\end{acknowledgements}

\bibliographystyle{aa} % style aa.bst
\bibliography{doar44_vlti} % your references ref1.bib

\begin{thebibliography}{48}
\expandafter\ifx\csname natexlab\endcsname\relax\def\natexlab#1{#1}\fi

\bibitem[{{Alencar} {et~al.}(2018){Alencar}, {Bouvier}, {Donati}, {Alecian},
  {Folsom}, {Grankin}, {Hussain}, {Hill}, {Cody}, {Carmona}, {Dougados},
  {Gregory}, {Herczeg}, {M{\'e}nard}, {Moutou}, {Malo}, {Takami}, \& {Matysse
  Collaboration}}]{Alencar18}
{Alencar}, S.~H.~P., {Bouvier}, J., {Donati}, J.~F., {et~al.} 2018, \aap, 620,
  A195

\bibitem[{{Alencar} {et~al.}(2012){Alencar}, {Bouvier}, {Walter}, {Dougados},
  {Donati}, {Kurosawa}, {Romanova}, {Bonfils}, {Lima}, {Massaro}, {Ibrahimov},
  \& {Poretti}}]{Alencar12}
{Alencar}, S.~H.~P., {Bouvier}, J., {Walter}, F.~M., {et~al.} 2012, \aap, 541,
  A116

\bibitem[{{Andrews} {et~al.}(2011){Andrews}, {Wilner}, {Espaillat}, {Hughes},
  {Dullemond}, {McClure}, {Qi}, \& {Brown}}]{Andrews11}
{Andrews}, S.~M., {Wilner}, D.~J., {Espaillat}, C., {et~al.} 2011, \apj, 732,
  42

\bibitem[{{Andrews} {et~al.}(2009){Andrews}, {Wilner}, {Hughes}, {Qi}, \&
  {Dullemond}}]{Andrews09}
{Andrews}, S.~M., {Wilner}, D.~J., {Hughes}, A.~M., {Qi}, C., \& {Dullemond},
  C.~P. 2009, \apj, 700, 1502

\bibitem[{{Anthonioz} {et~al.}(2015){Anthonioz}, {M{\'e}nard}, {Pinte}, {Le
  Bouquin}, {Benisty}, {Thi}, {Absil}, {Duch{\^e}ne}, {Augereau}, {Berger},
  {Casassus}, {Duvert}, {Lazareff}, {Malbet}, {Millan-Gabet}, {Schreiber},
  {Traub}, \& {Zins}}]{Anthonioz15}
{Anthonioz}, F., {M{\'e}nard}, F., {Pinte}, C., {et~al.} 2015, \aap, 574, A41

\bibitem[{{Bouvier} {et~al.}(2007{\natexlab{a}}){Bouvier}, {Alencar},
  {Boutelier}, {Dougados}, {Balog}, {Grankin}, {Hodgkin}, {Ibrahimov}, {Kun},
  {Magakian}, \& {Pinte}}]{Bouvier07a}
{Bouvier}, J., {Alencar}, S.~H.~P., {Boutelier}, T., {et~al.}
  2007{\natexlab{a}}, \aap, 463, 1017

\bibitem[{{Bouvier} {et~al.}(2007{\natexlab{b}}){Bouvier}, {Alencar},
  {Harries}, {Johns-Krull}, \& {Romanova}}]{Bouvier07b}
{Bouvier}, J., {Alencar}, S.~H.~P., {Harries}, T.~J., {Johns-Krull}, C.~M., \&
  {Romanova}, M.~M. 2007{\natexlab{b}}, in Protostars and Planets V, ed.
  B.~{Reipurth}, D.~{Jewitt}, \& K.~{Keil}, 479

\bibitem[{{Bouvier} \& {Appenzeller}(1992)}]{Bouvier92}
{Bouvier}, J. \& {Appenzeller}, I. 1992, \aaps, 92, 481

\bibitem[{{Casassus} {et~al.}(2018){Casassus}, {Avenhaus}, {P{\'e}rez},
  {Navarro}, {C{\'a}rcamo}, {Marino}, {Cieza}, {Quanz}, {Alarc{\'o}n}, {Zurlo},
  {Osses}, {Rannou}, {Rom{\'a}n}, \& {Barraza}}]{Casassus18}
{Casassus}, S., {Avenhaus}, H., {P{\'e}rez}, S., {et~al.} 2018, \mnras, 477,
  5104

\bibitem[{{Donati} {et~al.}(2019){Donati}, {Bouvier}, {Alencar}, {Hill},
  {Carmona}, {Folsom}, {M{\'e}nard}, {Gregory}, {Hussain}, {Grankin}, {Moutou},
  {Malo}, {Takami}, {Herczeg}, \& {MaTYSSE Collaboration}}]{Donati19}
{Donati}, J.~F., {Bouvier}, J., {Alencar}, S.~H., {et~al.} 2019, \mnras, 483,
  L1

\bibitem[{{Donati} {et~al.}(2020){Donati}, {Bouvier}, {Alencar}, {Moutou},
  {Malo}, {Takami}, {M{\'e}nard}, {Dougados}, {Hussain}, \& {The Matysse
  Collaboration}}]{Donati20}
{Donati}, J.~F., {Bouvier}, J., {Alencar}, S.~H., {et~al.} 2020, \mnras, 491,
  5660

\bibitem[{{Donati} {et~al.}(2013){Donati}, {Gregory}, {Alencar}, {Hussain},
  {Bouvier}, {Jardine}, {M{\'e}nard}, {Dougados}, {Romanova}, \& {MaPP
  Collaboration}}]{Donati13}
{Donati}, J.~F., {Gregory}, S.~G., {Alencar}, S.~H.~P., {et~al.} 2013, \mnras,
  436, 881

\bibitem[{{Eisner} {et~al.}(2009){Eisner}, {Graham}, {Akeson}, \&
  {Najita}}]{Eisner09}
{Eisner}, J.~A., {Graham}, J.~R., {Akeson}, R.~L., \& {Najita}, J. 2009, \apj,
  692, 309

\bibitem[{{Eisner} {et~al.}(2014){Eisner}, {Hillenbrand}, \&
  {Stone}}]{Eisner14}
{Eisner}, J.~A., {Hillenbrand}, L.~A., \& {Stone}, J.~M. 2014, \mnras, 443,
  1916

\bibitem[{{Eisner} {et~al.}(2010){Eisner}, {Monnier}, {Woillez}, {Akeson},
  {Millan-Gabet}, {Graham}, {Hillenbrand}, {Pott}, {Ragland}, \&
  {Wizinowich}}]{Eisner10}
{Eisner}, J.~A., {Monnier}, J.~D., {Woillez}, J., {et~al.} 2010, \apj, 718, 774

\bibitem[{{Espaillat} {et~al.}(2010){Espaillat}, {D'Alessio}, {Hern{\'a}ndez},
  {Nagel}, {Luhman}, {Watson}, {Calvet}, {Muzerolle}, \&
  {McClure}}]{Espaillat10}
{Espaillat}, C., {D'Alessio}, P., {Hern{\'a}ndez}, J., {et~al.} 2010, \apj,
  717, 441

\bibitem[{{Gaia Collaboration} {et~al.}(2018){Gaia Collaboration}, {Brown},
  {Vallenari}, {Prusti}, {de Bruijne}, {Babusiaux}, {Bailer-Jones}, {Biermann},
  {Evans}, {Eyer}, {Jansen}, {Jordi}, {Klioner}, {Lammers}, {Lindegren},
  {Luri}, {Mignard}, {Panem}, {Pourbaix}, {Randich}, {Sartoretti}, {Siddiqui},
  {Soubiran}, {van Leeuwen}, {Walton}, {Arenou}, {Bastian}, {Cropper},
  {Drimmel}, {Katz}, {Lattanzi}, {Bakker}, {Cacciari}, {Casta{\~n}eda},
  {Chaoul}, {Cheek}, {De Angeli}, {Fabricius}, {Guerra}, {Holl}, {Masana},
  {Messineo}, {Mowlavi}, {Nienartowicz}, {Panuzzo}, {Portell}, {Riello},
  {Seabroke}, {Tanga}, {Th{\'e}venin}, {Gracia-Abril}, {Comoretto},
  {Garcia-Reinaldos}, {Teyssier}, {Altmann}, {Andrae}, {Audard},
  {Bellas-Velidis}, {Benson}, {Berthier}, {Blomme}, {Burgess}, {Busso},
  {Carry}, {Cellino}, {Clementini}, {Clotet}, {Creevey}, {Davidson}, {De
  Ridder}, {Delchambre}, {Dell'Oro}, {Ducourant},
  {Fern{\'a}ndez-Hern{\'a}ndez}, {Fouesneau}, {Fr{\'e}mat}, {Galluccio},
  {Garc{\'\i}a-Torres}, {Gonz{\'a}lez-N{\'u}{\~n}ez}, {Gonz{\'a}lez-Vidal},
  {Gosset}, {Guy}, {Halbwachs}, {Hambly}, {Harrison}, {Hern{\'a}ndez},
  {Hestroffer}, {Hodgkin}, {Hutton}, {Jasniewicz}, {Jean-Antoine-Piccolo},
  {Jordan}, {Korn}, {Krone-Martins}, {Lanzafame}, {Lebzelter}, {L{\"o}ffler},
  {Manteiga}, {Marrese}, {Mart{\'\i}n-Fleitas}, {Moitinho}, {Mora}, {Muinonen},
  {Osinde}, {Pancino}, {Pauwels}, {Petit}, {Recio-Blanco}, {Richards},
  {Rimoldini}, {Robin}, {Sarro}, {Siopis}, {Smith}, {Sozzetti}, {S{\"u}veges},
  {Torra}, {van Reeven}, {Abbas}, {Abreu Aramburu}, {Accart}, {Aerts},
  {Altavilla}, {{\'A}lvarez}, {Alvarez}, {Alves}, {Anderson}, {Andrei},
  {Anglada Varela}, {Antiche}, {Antoja}, {Arcay}, {Astraatmadja}, {Bach},
  {Baker}, {Balaguer-N{\'u}{\~n}ez}, {Balm}, {Barache}, {Barata}, {Barbato},
  {Barblan}, {Barklem}, {Barrado}, {Barros}, {Barstow}, {Bartholom{\'e}
  Mu{\~n}oz}, {Bassilana}, {Becciani}, {Bellazzini}, {Berihuete}, {Bertone},
  {Bianchi}, {Bienaym{\'e}}, {Blanco-Cuaresma}, {Boch}, {Boeche}, {Bombrun},
  {Borrachero}, {Bossini}, {Bouquillon}, {Bourda}, {Bragaglia}, {Bramante},
  {Breddels}, {Bressan}, {Brouillet}, {Br{\"u}semeister}, {Brugaletta},
  {Bucciarelli}, {Burlacu}, {Busonero}, {Butkevich}, {Buzzi}, {Caffau},
  {Cancelliere}, {Cannizzaro}, {Cantat-Gaudin}, {Carballo}, {Carlucci},
  {Carrasco}, {Casamiquela}, {Castellani}, {Castro-Ginard}, {Charlot},
  {Chemin}, {Chiavassa}, {Cocozza}, {Costigan}, {Cowell}, {Crifo}, {Crosta},
  {Crowley}, {Cuypers}, {Dafonte}, {Damerdji}, {Dapergolas}, {David}, {David},
  {de Laverny}, {De Luise}, {De March}, {de Martino}, {de Souza}, {de Torres},
  {Debosscher}, {del Pozo}, {Delbo}, {Delgado}, {Delgado}, {Di Matteo},
  {Diakite}, {Diener}, {Distefano}, {Dolding}, {Drazinos}, {Dur{\'a}n},
  {Edvardsson}, {Enke}, {Eriksson}, {Esquej}, {Eynard Bontemps}, {Fabre},
  {Fabrizio}, {Faigler}, {Falc{\~a}o}, {Farr{\`a}s Casas}, {Federici},
  {Fedorets}, {Fernique}, {Figueras}, {Filippi}, {Findeisen}, {Fonti},
  {Fraile}, {Fraser}, {Fr{\'e}zouls}, {Gai}, {Galleti}, {Garabato},
  {Garc{\'\i}a-Sedano}, {Garofalo}, {Garralda}, {Gavel}, {Gavras}, {Gerssen},
  {Geyer}, {Giacobbe}, {Gilmore}, {Girona}, {Giuffrida}, {Glass}, {Gomes},
  {Granvik}, {Gueguen}, {Guerrier}, {Guiraud}, {Guti{\'e}rrez-S{\'a}nchez},
  {Haigron}, {Hatzidimitriou}, {Hauser}, {Haywood}, {Heiter}, {Helmi}, {Heu},
  {Hilger}, {Hobbs}, {Hofmann}, {Holland}, {Huckle}, {Hypki}, {Icardi},
  {Jan{\ss}en}, {Jevardat de Fombelle}, {Jonker}, {Juh{\'a}sz}, {Julbe},
  {Karampelas}, {Kewley}, {Klar}, {Kochoska}, {Kohley}, {Kolenberg},
  {Kontizas}, {Kontizas}, {Koposov}, {Kordopatis}, {Kostrzewa-Rutkowska},
  {Koubsky}, {Lambert}, {Lanza}, {Lasne}, {Lavigne}, {Le Fustec}, {Le
  Poncin-Lafitte}, {Lebreton}, {Leccia}, {Leclerc}, {Lecoeur-Taibi},
  {Lenhardt}, {Leroux}, {Liao}, {Licata}, {Lindstr{\o}m}, {Lister}, {Livanou},
  {Lobel}, {L{\'o}pez}, {Managau}, {Mann}, {Mantelet}, {Marchal}, {Marchant},
  {Marconi}, {Marinoni}, {Marschalk{\'o}}, {Marshall}, {Martino}, {Marton},
  {Mary}, {Massari}, {Matijevi{\v{c}}}, {Mazeh}, {McMillan}, {Messina},
  {Michalik}, {Millar}, {Molina}, {Molinaro}, {Moln{\'a}r}, {Montegriffo},
  {Mor}, {Morbidelli}, {Morel}, {Morris}, {Mulone}, {Muraveva}, {Musella},
  {Nelemans}, {Nicastro}, {Noval}, {O'Mullane}, {Ord{\'e}novic},
  {Ord{\'o}{\~n}ez-Blanco}, {Osborne}, {Pagani}, {Pagano}, {Pailler},
  {Palacin}, {Palaversa}, {Panahi}, {Pawlak}, {Piersimoni}, {Pineau}, {Plachy},
  {Plum}, {Poggio}, {Poujoulet}, {Pr{\v{s}}a}, {Pulone}, {Racero}, {Ragaini},
  {Rambaux}, {Ramos-Lerate}, {Regibo}, {Reyl{\'e}}, {Riclet}, {Ripepi}, {Riva},
  {Rivard}, {Rixon}, {Roegiers}, {Roelens}, {Romero-G{\'o}mez}, {Rowell},
  {Royer}, {Ruiz-Dern}, {Sadowski}, {Sagrist{\`a} Sell{\'e}s}, {Sahlmann},
  {Salgado}, {Salguero}, {Sanna}, {Santana-Ros}, {Sarasso}, {Savietto},
  {Schultheis}, {Sciacca}, {Segol}, {Segovia}, {S{\'e}gransan}, {Shih},
  {Siltala}, {Silva}, {Smart}, {Smith}, {Solano}, {Solitro}, {Sordo}, {Soria
  Nieto}, {Souchay}, {Spagna}, {Spoto}, {Stampa}, {Steele},
  {Steidelm{\"u}ller}, {Stephenson}, {Stoev}, {Suess}, {Surdej}, {Szabados},
  {Szegedi-Elek}, {Tapiador}, {Taris}, {Tauran}, {Taylor}, {Teixeira},
  {Terrett}, {Teyssand ier}, {Thuillot}, {Titarenko}, {Torra Clotet}, {Turon},
  {Ulla}, {Utrilla}, {Uzzi}, {Vaillant}, {Valentini}, {Valette}, {van Elteren},
  {Van Hemelryck}, {van Leeuwen}, {Vaschetto}, {Vecchiato}, {Veljanoski},
  {Viala}, {Vicente}, {Vogt}, {von Essen}, {Voss}, {Votruba}, {Voutsinas},
  {Walmsley}, {Weiler}, {Wertz}, {Wevers}, {Wyrzykowski}, {Yoldas},
  {{\v{Z}}erjal}, {Ziaeepour}, {Zorec}, {Zschocke}, {Zucker}, {Zurbach}, \&
  {Zwitter}}]{Gaia18}
{Gaia Collaboration}, {Brown}, A.~G.~A., {Vallenari}, A., {et~al.} 2018, \aap,
  616, A1

\bibitem[{{Gaia Collaboration} {et~al.}(2016){Gaia Collaboration}, {Prusti},
  {de Bruijne}, {Brown}, {Vallenari}, {Babusiaux}, {Bailer-Jones}, {Bastian},
  {Biermann}, {Evans}, {Eyer}, {Jansen}, {Jordi}, {Klioner}, {Lammers},
  {Lindegren}, {Luri}, {Mignard}, {Milligan}, {Panem}, {Poinsignon},
  {Pourbaix}, {Randich}, {Sarri}, {Sartoretti}, {Siddiqui}, {Soubiran},
  {Valette}, {van Leeuwen}, {Walton}, {Aerts}, {Arenou}, {Cropper}, {Drimmel},
  {H{\o}g}, {Katz}, {Lattanzi}, {O'Mullane}, {Grebel}, {Holland}, {Huc},
  {Passot}, {Bramante}, {Cacciari}, {Casta{\~n}eda}, {Chaoul}, {Cheek}, {De
  Angeli}, {Fabricius}, {Guerra}, {Hern{\'a}ndez}, {Jean-Antoine-Piccolo},
  {Masana}, {Messineo}, {Mowlavi}, {Nienartowicz}, {Ord{\'o}{\~n}ez-Blanco},
  {Panuzzo}, {Portell}, {Richards}, {Riello}, {Seabroke}, {Tanga},
  {Th{\'e}venin}, {Torra}, {Els}, {Gracia-Abril}, {Comoretto},
  {Garcia-Reinaldos}, {Lock}, {Mercier}, {Altmann}, {Andrae}, {Astraatmadja},
  {Bellas-Velidis}, {Benson}, {Berthier}, {Blomme}, {Busso}, {Carry},
  {Cellino}, {Clementini}, {Cowell}, {Creevey}, {Cuypers}, {Davidson}, {De
  Ridder}, {de Torres}, {Delchambre}, {Dell'Oro}, {Ducourant}, {Fr{\'e}mat},
  {Garc{\'\i}a-Torres}, {Gosset}, {Halbwachs}, {Hambly}, {Harrison}, {Hauser},
  {Hestroffer}, {Hodgkin}, {Huckle}, {Hutton}, {Jasniewicz}, {Jordan},
  {Kontizas}, {Korn}, {Lanzafame}, {Manteiga}, {Moitinho}, {Muinonen},
  {Osinde}, {Pancino}, {Pauwels}, {Petit}, {Recio-Blanco}, {Robin}, {Sarro},
  {Siopis}, {Smith}, {Smith}, {Sozzetti}, {Thuillot}, {van Reeven}, {Viala},
  {Abbas}, {Abreu Aramburu}, {Accart}, {Aguado}, {Allan}, {Allasia},
  {Altavilla}, {{\'A}lvarez}, {Alves}, {Anderson}, {Andrei}, {Anglada Varela},
  {Antiche}, {Antoja}, {Ant{\'o}n}, {Arcay}, {Atzei}, {Ayache}, {Bach},
  {Baker}, {Balaguer-N{\'u}{\~n}ez}, {Barache}, {Barata}, {Barbier}, {Barblan},
  {Baroni}, {Barrado y Navascu{\'e}s}, {Barros}, {Barstow}, {Becciani},
  {Bellazzini}, {Bellei}, {Bello Garc{\'\i}a}, {Belokurov}, {Bendjoya},
  {Berihuete}, {Bianchi}, {Bienaym{\'e}}, {Billebaud}, {Blagorodnova},
  {Blanco-Cuaresma}, {Boch}, {Bombrun}, {Borrachero}, {Bouquillon}, {Bourda},
  {Bouy}, {Bragaglia}, {Breddels}, {Brouillet}, {Br{\"u}semeister},
  {Bucciarelli}, {Budnik}, {Burgess}, {Burgon}, {Burlacu}, {Busonero}, {Buzzi},
  {Caffau}, {Cambras}, {Campbell}, {Cancelliere}, {Cantat-Gaudin}, {Carlucci},
  {Carrasco}, {Castellani}, {Charlot}, {Charnas}, {Charvet}, {Chassat},
  {Chiavassa}, {Clotet}, {Cocozza}, {Collins}, {Collins}, {Costigan}, {Crifo},
  {Cross}, {Crosta}, {Crowley}, {Dafonte}, {Damerdji}, {Dapergolas}, {David},
  {David}, {De Cat}, {de Felice}, {de Laverny}, {De Luise}, {De March}, {de
  Martino}, {de Souza}, {Debosscher}, {del Pozo}, {Delbo}, {Delgado},
  {Delgado}, {di Marco}, {Di Matteo}, {Diakite}, {Distefano}, {Dolding}, {Dos
  Anjos}, {Drazinos}, {Dur{\'a}n}, {Dzigan}, {Ecale}, {Edvardsson}, {Enke},
  {Erdmann}, {Escolar}, {Espina}, {Evans}, {Eynard Bontemps}, {Fabre},
  {Fabrizio}, {Faigler}, {Falc{\~a}o}, {Farr{\`a}s Casas}, {Faye}, {Federici},
  {Fedorets}, {Fern{\'a}ndez-Hern{\'a}ndez}, {Fernique}, {Fienga}, {Figueras},
  {Filippi}, {Findeisen}, {Fonti}, {Fouesneau}, {Fraile}, {Fraser}, {Fuchs},
  {Furnell}, {Gai}, {Galleti}, {Galluccio}, {Garabato}, {Garc{\'\i}a-Sedano},
  {Gar{\'e}}, {Garofalo}, {Garralda}, {Gavras}, {Gerssen}, {Geyer}, {Gilmore},
  {Girona}, {Giuffrida}, {Gomes}, {Gonz{\'a}lez-Marcos},
  {Gonz{\'a}lez-N{\'u}{\~n}ez}, {Gonz{\'a}lez-Vidal}, {Granvik}, {Guerrier},
  {Guillout}, {Guiraud}, {G{\'u}rpide}, {Guti{\'e}rrez-S{\'a}nchez}, {Guy},
  {Haigron}, {Hatzidimitriou}, {Haywood}, {Heiter}, {Helmi}, {Hobbs},
  {Hofmann}, {Holl}, {Holland }, {Hunt}, {Hypki}, {Icardi}, {Irwin}, {Jevardat
  de Fombelle}, {Jofr{\'e}}, {Jonker}, {Jorissen}, {Julbe}, {Karampelas},
  {Kochoska}, {Kohley}, {Kolenberg}, {Kontizas}, {Koposov}, {Kordopatis},
  {Koubsky}, {Kowalczyk}, {Krone-Martins}, {Kudryashova}, {Kull}, {Bachchan},
  {Lacoste-Seris}, {Lanza}, {Lavigne}, {Le Poncin-Lafitte}, {Lebreton},
  {Lebzelter}, {Leccia}, {Leclerc}, {Lecoeur-Taibi}, {Lemaitre}, {Lenhardt},
  {Leroux}, {Liao}, {Licata}, {Lindstr{\o}m}, {Lister}, {Livanou}, {Lobel},
  {L{\"o}ffler}, {L{\'o}pez}, {Lopez-Lozano}, {Lorenz}, {Loureiro},
  {MacDonald}, {Magalh{\~a}es Fernandes}, {Managau}, {Mann}, {Mantelet},
  {Marchal}, {Marchant}, {Marconi}, {Marie}, {Marinoni}, {Marrese},
  {Marschalk{\'o}}, {Marshall}, {Mart{\'\i}n-Fleitas}, {Martino}, {Mary},
  {Matijevi{\v{c}}}, {Mazeh}, {McMillan}, {Messina}, {Mestre}, {Michalik},
  {Millar}, {Miranda}, {Molina}, {Molinaro}, {Molinaro}, {Moln{\'a}r},
  {Moniez}, {Montegriffo}, {Monteiro}, {Mor}, {Mora}, {Morbidelli}, {Morel},
  {Morgenthaler}, {Morley}, {Morris}, {Mulone}, {Muraveva}, {Musella},
  {Narbonne}, {Nelemans}, {Nicastro}, {Noval}, {Ord{\'e}novic},
  {Ordieres-Mer{\'e}}, {Osborne}, {Pagani}, {Pagano}, {Pailler}, {Palacin},
  {Palaversa}, {Parsons}, {Paulsen}, {Pecoraro}, {Pedrosa}, {Pentik{\"a}inen},
  {Pereira}, {Pichon}, {Piersimoni}, {Pineau}, {Plachy}, {Plum}, {Poujoulet},
  {Pr{\v{s}}a}, {Pulone}, {Ragaini}, {Rago}, {Rambaux}, {Ramos-Lerate},
  {Ranalli}, {Rauw}, {Read}, {Regibo}, {Renk}, {Reyl{\'e}}, {Ribeiro},
  {Rimoldini}, {Ripepi}, {Riva}, {Rixon}, {Roelens}, {Romero-G{\'o}mez},
  {Rowell}, {Royer}, {Rudolph}, {Ruiz-Dern}, {Sadowski}, {Sagrist{\`a}
  Sell{\'e}s}, {Sahlmann}, {Salgado}, {Salguero}, {Sarasso}, {Savietto},
  {Schnorhk}, {Schultheis}, {Sciacca}, {Segol}, {Segovia}, {Segransan},
  {Serpell}, {Shih}, {Smareglia}, {Smart}, {Smith}, {Solano}, {Solitro},
  {Sordo}, {Soria Nieto}, {Souchay}, {Spagna}, {Spoto}, {Stampa}, {Steele},
  {Steidelm{\"u}ller}, {Stephenson}, {Stoev}, {Suess}, {S{\"u}veges}, {Surdej},
  {Szabados}, {Szegedi-Elek}, {Tapiador}, {Taris}, {Tauran}, {Taylor},
  {Teixeira}, {Terrett}, {Tingley}, {Trager}, {Turon}, {Ulla}, {Utrilla},
  {Valentini}, {van Elteren}, {Van Hemelryck}, {van Leeuwen}, {Varadi},
  {Vecchiato}, {Veljanoski}, {Via}, {Vicente}, {Vogt}, {Voss}, {Votruba},
  {Voutsinas}, {Walmsley}, {Weiler}, {Weingrill}, {Werner}, {Wevers},
  {Whitehead}, {Wyrzykowski}, {Yoldas}, {{\v{Z}}erjal}, {Zucker}, {Zurbach},
  {Zwitter}, {Alecu}, {Allen}, {Allende Prieto}, {Amorim},
  {Anglada-Escud{\'e}}, {Arsenijevic}, {Azaz}, {Balm}, {Beck}, {Bernstein},
  {Bigot}, {Bijaoui}, {Blasco}, {Bonfigli}, {Bono}, {Boudreault}, {Bressan},
  {Brown}, {Brunet}, {Bunclark}, {Buonanno}, {Butkevich}, {Carret}, {Carrion},
  {Chemin}, {Ch{\'e}reau}, {Corcione}, {Darmigny}, {de Boer}, {de Teodoro}, {de
  Zeeuw}, {Delle Luche}, {Domingues}, {Dubath}, {Fodor}, {Fr{\'e}zouls},
  {Fries}, {Fustes}, {Fyfe}, {Gallardo}, {Gallegos}, {Gardiol}, {Gebran},
  {Gomboc}, {G{\'o}mez}, {Grux}, {Gueguen}, {Heyrovsky}, {Hoar}, {Iannicola},
  {Isasi Parache}, {Janotto}, {Joliet}, {Jonckheere}, {Keil}, {Kim},
  {Klagyivik}, {Klar}, {Knude}, {Kochukhov}, {Kolka}, {Kos}, {Kutka}, {Lainey},
  {LeBouquin}, {Liu}, {Loreggia}, {Makarov}, {Marseille}, {Martayan},
  {Martinez-Rubi}, {Massart}, {Meynadier}, {Mignot}, {Munari}, {Nguyen},
  {Nordlander}, {Ocvirk}, {O'Flaherty}, {Olias Sanz}, {Ortiz}, {Osorio},
  {Oszkiewicz}, {Ouzounis}, {Palmer}, {Park}, {Pasquato}, {Peltzer}, {Peralta},
  {P{\'e}turaud}, {Pieniluoma}, {Pigozzi}, {Poels}, {Prat}, {Prod'homme},
  {Raison}, {Rebordao}, {Risquez}, {Rocca-Volmerange}, {Rosen}, {Ruiz-Fuertes},
  {Russo}, {Sembay}, {Serraller Vizcaino}, {Short}, {Siebert}, {Silva},
  {Sinachopoulos}, {Slezak}, {Soffel}, {Sosnowska}, {Strai{\v{z}}ys}, {ter
  Linden}, {Terrell}, {Theil}, {Tiede}, {Troisi}, {Tsalmantza}, {Tur},
  {Vaccari}, {Vachier}, {Valles}, {Van Hamme}, {Veltz}, {Virtanen}, {Wallut},
  {Wichmann}, {Wilkinson}, {Ziaeepour}, \& {Zschocke}}]{Gaia16}
{Gaia Collaboration}, {Prusti}, T., {de Bruijne}, J.~H.~J., {et~al.} 2016,
  \aap, 595, A1

\bibitem[{{Goto} {et~al.}(2012){Goto}, {Carmona}, {Linz}, {Stecklum},
  {Henning}, {Meeus}, \& {Usuda}}]{Goto12}
{Goto}, M., {Carmona}, A., {Linz}, H., {et~al.} 2012, \apj, 748, 6

\bibitem[{{Gravity Collaboration} {et~al.}(2017{\natexlab{a}}){Gravity
  Collaboration}, {Abuter}, {Accardo}, {Amorim}, {Anugu}, {{\'A}vila},
  {Azouaoui}, {Benisty}, {Berger}, {Blind}, {Bonnet}, {Bourget}, {Brandner},
  {Brast}, {Buron}, {Burtscher}, {Cassaing}, {Chapron}, {Choquet},
  {Cl{\'e}net}, {Collin}, {Coud{\'e} Du Foresto}, {de Wit}, {de Zeeuw}, {Deen},
  {Delplancke-Str{\"o}bele}, {Dembet}, {Derie}, {Dexter}, {Duvert}, {Ebert},
  {Eckart}, {Eisenhauer}, {Esselborn}, {F{\'e}dou}, {Finger}, {Garcia}, {Garcia
  Dabo}, {Garcia Lopez}, {Gendron}, {Genzel}, {Gillessen}, {Gonte}, {Gordo},
  {Grould}, {Gr{\"o}zinger}, {Guieu}, {Haguenauer}, {Hans}, {Haubois}, {Haug},
  {Haussmann}, {Henning}, {Hippler}, {Horrobin}, {Huber}, {Hubert}, {Hubin},
  {Hummel}, {Jakob}, {Janssen}, {Jochum}, {Jocou}, {Kaufer}, {Kellner},
  {Kendrew}, {Kern}, {Kervella}, {Kiekebusch}, {Klein}, {Kok}, {Kolb}, {Kulas},
  {Lacour}, {Lapeyr{\`e}re}, {Lazareff}, {Le Bouquin}, {L{\`e}na}, {Lenzen},
  {L{\'e}v{\^e}que}, {Lippa}, {Magnard}, {Mehrgan}, {Mellein}, {M{\'e}rand},
  {Moreno-Ventas}, {Moulin}, {M{\"u}ller}, {M{\"u}ller}, {Neumann}, {Oberti},
  {Ott}, {Pallanca}, {Panduro}, {Pasquini}, {Paumard}, {Percheron}, {Perraut},
  {Perrin}, {Pfl{\"u}ger}, {Pfuhl}, {Phan Duc}, {Plewa}, {Popovic}, {Rabien},
  {Ram{\'\i}rez}, {Ramos}, {Rau}, {Riquelme}, {Rohloff}, {Rousset},
  {Sanchez-Bermudez}, {Scheithauer}, {Sch{\"o}ller}, {Schuhler}, {Spyromilio},
  {Straubmeier}, {Sturm}, {Suarez}, {Tristram}, {Ventura}, {Vincent},
  {Waisberg}, {Wank}, {Weber}, {Wieprecht}, {Wiest}, {Wiezorrek}, {Wittkowski},
  {Woillez}, {Wolff}, {Yazici}, {Ziegler}, \& {Zins}}]{Grav17}
{Gravity Collaboration}, {Abuter}, R., {Accardo}, M., {et~al.}
  2017{\natexlab{a}}, \aap, 602, A94

\bibitem[{{Gravity Collaboration} {et~al.}(2020){Gravity Collaboration},
  {Caratti o Garatti}, {Fedriani}, {Garcia Lopez}, {Koutoulaki}, \&
  {Perraut}}]{Gravity20}
{Gravity Collaboration}, {Caratti o Garatti}, A., {Fedriani}, R., {et~al.}
  2020, https://arxiv.org/abs/2003.05404

\bibitem[{{Gravity Collaboration} {et~al.}(2017{\natexlab{b}}){Gravity
  Collaboration}, {Garcia Lopez}, {Perraut}, {Caratti O Garatti}, {Lazareff},
  {Sanchez-Bermudez}, {Benisty}, {Dougados}, {Labadie}, {Brandner}, {Garcia},
  {Henning}, {Ray}, {Abuter}, {Amorim}, {Anugu}, {Berger}, {Bonnet}, {Buron},
  {Caselli}, {Cl{\'e}net}, {Coud{\'e} Du Foresto}, {de Wit}, {Deen},
  {Delplancke-Str{\"o}bele}, {Dexter}, {Eckart}, {Eisenhauer}, {Garcia Dabo},
  {Gendron}, {Genzel}, {Gillessen}, {Haubois}, {Haug}, {Haussmann}, {Hippler},
  {Hubert}, {Hummel}, {Horrobin}, {Jocou}, {Kellner}, {Kervella}, {Kulas},
  {Kolb}, {Lacour}, {Le Bouquin}, {L{\'e}na}, {Lippa}, {M{\'e}rand},
  {M{\"u}ller}, {Ott}, {Panduro}, {Paumard}, {Perrin}, {Pfuhl}, {Ramirez},
  {Rau}, {Rohloff}, {Rousset}, {Scheithauer}, {Sch{\"o}ller}, {Straubmeier},
  {Sturm}, {Thi}, {van Dishoeck}, {Vincent}, {Waisberg}, {Wank}, {Wieprecht},
  {Wiest}, {Wiezorrek}, {Woillez}, {Yazici}, \& {Zins}}]{Gravity17}
{Gravity Collaboration}, {Garcia Lopez}, R., {Perraut}, K., {et~al.}
  2017{\natexlab{b}}, \aap, 608, A78

\bibitem[{{Gravity Collaboration} {et~al.}(2019){Gravity Collaboration},
  {Perraut}, {Labadie}, {Lazareff}, {Klarmann}, {Segura-Cox}, {Benisty},
  {Bouvier}, {Brandner}, {Caratti O Garatti}, {Caselli}, {Dougados}, {Garcia},
  {Garcia-Lopez}, {Kendrew}, {Koutoulaki}, {Kervella}, {Lin}, {Pineda},
  {Sanchez-Bermudez}, {van Dishoeck}, {Abuter}, {Amorim}, {Berger}, {Bonnet},
  {Buron}, {Cantalloube}, {Cl{\'e}net}, {Coud{\'e} Du Foresto}, {Dexter}, {de
  Zeeuw}, {Duvert}, {Eckart}, {Eisenhauer}, {Eupen}, {Gao}, {Gendron},
  {Genzel}, {Gillessen}, {Gordo}, {Grellmann}, {Haubois}, {Haussmann},
  {Henning}, {Hippler}, {Horrobin}, {Hubert}, {Jocou}, {Lacour}, {Le Bouquin},
  {L{\'e}na}, {M{\'e}rand}, {Ott}, {Paumard}, {Perrin}, {Pfuhl}, {Rabien},
  {Ray}, {Rau}, {Rousset}, {Scheithauer}, {Straub}, {Straubmeier}, {Sturm},
  {Vincent}, {Waisberg}, {Wank}, {Widmann}, {Wieprecht}, {Wiest}, {Wiezorrek},
  {Woillez}, \& {Yazici}}]{Gravity19}
{Gravity Collaboration}, {Perraut}, K., {Labadie}, L., {et~al.} 2019, \aap,
  632, A53

\bibitem[{{Haffert} {et~al.}(2019){Haffert}, {Bohn}, {de Boer}, {Snellen},
  {Brinchmann}, {Girard}, {Keller}, \& {Bacon}}]{Haffert19}
{Haffert}, S.~Y., {Bohn}, A.~J., {de Boer}, J., {et~al.} 2019, Nature
  Astronomy, 3, 749

\bibitem[{{Hartmann} {et~al.}(2016){Hartmann}, {Herczeg}, \&
  {Calvet}}]{Hartmann16}
{Hartmann}, L., {Herczeg}, G., \& {Calvet}, N. 2016, \araa, 54, 135

\bibitem[{{Kraus} {et~al.}(2008){Kraus}, {Hofmann}, {Benisty}, {Berger},
  {Chesneau}, {Isella}, {Malbet}, {Meilland}, {Nardetto}, {Natta}, {Preibisch},
  {Schertl}, {Smith}, {Stee}, {Tatulli}, {Testi}, \& {Weigelt}}]{Kraus08}
{Kraus}, S., {Hofmann}, K.~H., {Benisty}, M., {et~al.} 2008, \aap, 489, 1157

\bibitem[{{Kreplin} {et~al.}(2018){Kreplin}, {Tambovtseva}, {Grinin}, {Kraus},
  {Weigelt}, \& {Wang}}]{Kreplin18}
{Kreplin}, A., {Tambovtseva}, L., {Grinin}, V., {et~al.} 2018, \mnras, 476,
  4520

\bibitem[{{Kurosawa} \& {Romanova}(2012)}]{Kurosawa12}
{Kurosawa}, R. \& {Romanova}, M.~M. 2012, \mnras, 426, 2901

\bibitem[{{Kurosawa} \& {Romanova}(2013)}]{Kurosawa13}
{Kurosawa}, R. \& {Romanova}, M.~M. 2013, \mnras, 431, 2673

\bibitem[{{Kurosawa} {et~al.}(2008){Kurosawa}, {Romanova}, \&
  {Harries}}]{Kurosawa08}
{Kurosawa}, R., {Romanova}, M.~M., \& {Harries}, T.~J. 2008, \mnras, 385, 1931

\bibitem[{{Kurosawa} {et~al.}(2011){Kurosawa}, {Romanova}, \&
  {Harries}}]{Kurosawa11}
{Kurosawa}, R., {Romanova}, M.~M., \& {Harries}, T.~J. 2011, \mnras, 416, 2623

\bibitem[{{Lacour} {et~al.}(2019){Lacour}, {Dembet}, {Abuter}, {F{\'e}dou},
  {Perrin}, {Choquet}, {Pfuhl}, {Eisenhauer}, {Woillez}, {Cassaing},
  {Wieprecht}, {Ott}, {Wiezorrek}, {Tristram}, {Wolff}, {Ram{\'\i}rez},
  {Haubois}, {Perraut}, {Straubmeier}, {Brand ner}, \& {Amorim}}]{Lacour19}
{Lacour}, S., {Dembet}, R., {Abuter}, R., {et~al.} 2019, \aap, 624, A99

\bibitem[{{Lapeyrere} {et~al.}(2014){Lapeyrere}, {Kervella}, {Lacour},
  {Azouaoui}, {Garcia-Dabo}, {Perrin}, {Eisenhauer}, {Perraut}, {Straubmeier},
  {Amorim}, \& {Brandner}}]{Lapeyrere14}
{Lapeyrere}, V., {Kervella}, P., {Lacour}, S., {et~al.} 2014, Society of
  Photo-Optical Instrumentation Engineers (SPIE) Conference Series, Vol. 9146,
  {GRAVITY data reduction software}, 91462D

\bibitem[{{Lavail} {et~al.}(2017){Lavail}, {Kochukhov}, {Hussain}, {Alecian},
  {Herczeg}, \& {Johns-Krull}}]{Lavail17}
{Lavail}, A., {Kochukhov}, O., {Hussain}, G.~A.~J., {et~al.} 2017, \aap, 608,
  A77

\bibitem[{{Lazareff} {et~al.}(2017){Lazareff}, {Berger}, {Kluska}, {Le
  Bouquin}, {Benisty}, {Malbet}, {Koen}, {Pinte}, {Thi}, {Absil}, {Baron},
  {Delboulb{\'e}}, {Duvert}, {Isella}, {Jocou}, {Juhasz}, {Kraus}, {Lachaume},
  {M{\'e}nard}, {Millan-Gabet}, {Monnier}, {Moulin}, {Perraut}, {Rochat},
  {Soulez}, {Tallon}, {Thi{\'e}baut}, {Traub}, \& {Zins}}]{Lazareff17}
{Lazareff}, B., {Berger}, J.~P., {Kluska}, J., {et~al.} 2017, \aap, 599, A85

\bibitem[{{Malbet} {et~al.}(2007){Malbet}, {Benisty}, {de Wit}, {Kraus},
  {Meilland}, {Millour}, {Tatulli}, {Berger}, {Chesneau}, {Hofmann}, {Isella},
  {Natta}, {Petrov}, {Preibisch}, {Stee}, {Testi}, {Weigelt}, {Antonelli},
  {Beckmann}, {Bresson}, {Chelli}, {Dugu{\'e}}, {Duvert}, {Gennari},
  {Gl{\"u}ck}, {Kern}, {Lagarde}, {Le Coarer}, {Lisi}, {Perraut}, {Puget},
  {Rantakyr{\"o}}, {Robbe-Dubois}, {Roussel}, {Zins}, {Accardo}, {Acke},
  {Agabi}, {Altariba}, {Arezki}, {Aristidi}, {Baffa}, {Behrend}, {Bl{\"o}cker},
  {Bonhomme}, {Busoni}, {Cassaing}, {Clausse}, {Colin}, {Connot},
  {Delboulb{\'e}}, {Domiciano de Souza}, {Driebe}, {Feautrier}, {Ferruzzi},
  {Forveille}, {Fossat}, {Foy}, {Fraix-Burnet}, {Gallardo}, {Giani}, {Gil},
  {Glentzlin}, {Heiden}, {Heininger}, {Hernandez Utrera}, {Kamm}, {Kiekebusch},
  {Le Contel}, {Le Contel}, {Lesourd}, {Lopez}, {Lopez}, {Magnard}, {Marconi},
  {Mars}, {Martinot-Lagarde}, {Mathias}, {M{\`e}ge}, {Monin}, {Mouillet},
  {Mourard}, {Nussbaum}, {Ohnaka}, {Pacheco}, {Perrier}, {Rabbia}, {Rebattu},
  {Reynaud}, {Richichi}, {Robini}, {Sacchettini}, {Schertl}, {Sch{\"o}ller},
  {Solscheid}, {Spang}, {Stefanini}, {Tallon}, {Tallon-Bosc}, {Tasso},
  {Vakili}, {von der L{\"u}he}, {Valtier}, {Vannier}, \& {Ventura}}]{Malbet07}
{Malbet}, F., {Benisty}, M., {de Wit}, W.~J., {et~al.} 2007, \aap, 464, 43

\bibitem[{{Manara} {et~al.}(2014){Manara}, {Testi}, {Natta}, {Rosotti},
  {Benisty}, {Ercolano}, \& {Ricci}}]{Manara14}
{Manara}, C.~F., {Testi}, L., {Natta}, A., {et~al.} 2014, \aap, 568, A18

\bibitem[{{Mendigut{\'\i}a} {et~al.}(2015){Mendigut{\'\i}a}, {de Wit},
  {Oudmaijer}, {Fairlamb}, {Carciofi}, {Ilee}, \& {Vieira}}]{Mendigutia15}
{Mendigut{\'\i}a}, I., {de Wit}, W.~J., {Oudmaijer}, R.~D., {et~al.} 2015,
  \mnras, 453, 2126

\bibitem[{{Monnier}(2007)}]{Monnier07}
{Monnier}, J.~D. 2007, \nar, 51, 604

\bibitem[{{Muzerolle} {et~al.}(1998{\natexlab{a}}){Muzerolle}, {Calvet}, \&
  {Hartmann}}]{Muzerolle98b}
{Muzerolle}, J., {Calvet}, N., \& {Hartmann}, L. 1998{\natexlab{a}}, \apj, 492,
  743

\bibitem[{{Muzerolle} {et~al.}(1998{\natexlab{b}}){Muzerolle}, {Hartmann}, \&
  {Calvet}}]{Muzerolle98a}
{Muzerolle}, J., {Hartmann}, L., \& {Calvet}, N. 1998{\natexlab{b}}, \aj, 116,
  2965

\bibitem[{{Najita} {et~al.}(1996){Najita}, {Carr}, \& {Tokunaga}}]{Najita96}
{Najita}, J., {Carr}, J.~S., \& {Tokunaga}, A.~T. 1996, \apj, 456, 292

\bibitem[{{Pinte} {et~al.}(2008){Pinte}, {M{\'e}nard}, {Berger}, {Benisty}, \&
  {Malbet}}]{Pinte2008}
{Pinte}, C., {M{\'e}nard}, F., {Berger}, J.~P., {Benisty}, M., \& {Malbet}, F.
  2008, \apjl, 673, L63

\bibitem[{{Salyk} {et~al.}(2009){Salyk}, {Blake}, {Boogert}, \&
  {Brown}}]{Salyk09}
{Salyk}, C., {Blake}, G.~A., {Boogert}, A.~C.~A., \& {Brown}, J.~M. 2009, \apj,
  699, 330

\bibitem[{{Salyk} {et~al.}(2011){Salyk}, {Blake}, {Boogert}, \&
  {Brown}}]{Salyk11}
{Salyk}, C., {Blake}, G.~A., {Boogert}, A.~C.~A., \& {Brown}, J.~M. 2011, \apj,
  743, 112

\bibitem[{{van der Marel} {et~al.}(2016){van der Marel}, {van Dishoeck},
  {Bruderer}, {Andrews}, {Pontoppidan}, {Herczeg}, {van Kempen}, \&
  {Miotello}}]{vanderMarel16}
{van der Marel}, N., {van Dishoeck}, E.~F., {Bruderer}, S., {et~al.} 2016,
  \aap, 585, A58

\bibitem[{{Weigelt} {et~al.}(2011){Weigelt}, {Grinin}, {Groh}, {Hofmann},
  {Kraus}, {Miroshnichenko}, {Schertl}, {Tambovtseva}, {Benisty}, {Driebe},
  {Lagarde}, {Malbet}, {Meilland}, {Petrov}, \& {Tatulli}}]{Weigelt11}
{Weigelt}, G., {Grinin}, V.~P., {Groh}, J.~H., {et~al.} 2011, \aap, 527, A103

\bibitem[{{Winn} {et~al.}(2018){Winn}, {Sanchis-Ojeda}, \&
  {Rappaport}}]{Winn18}
{Winn}, J.~N., {Sanchis-Ojeda}, R., \& {Rappaport}, S. 2018, \nar, 83, 37

\end{thebibliography}

\appendix

\section{Differential observables}

\begin{figure*}[h]
  \centering
  \includegraphics[width=0.48\hsize]{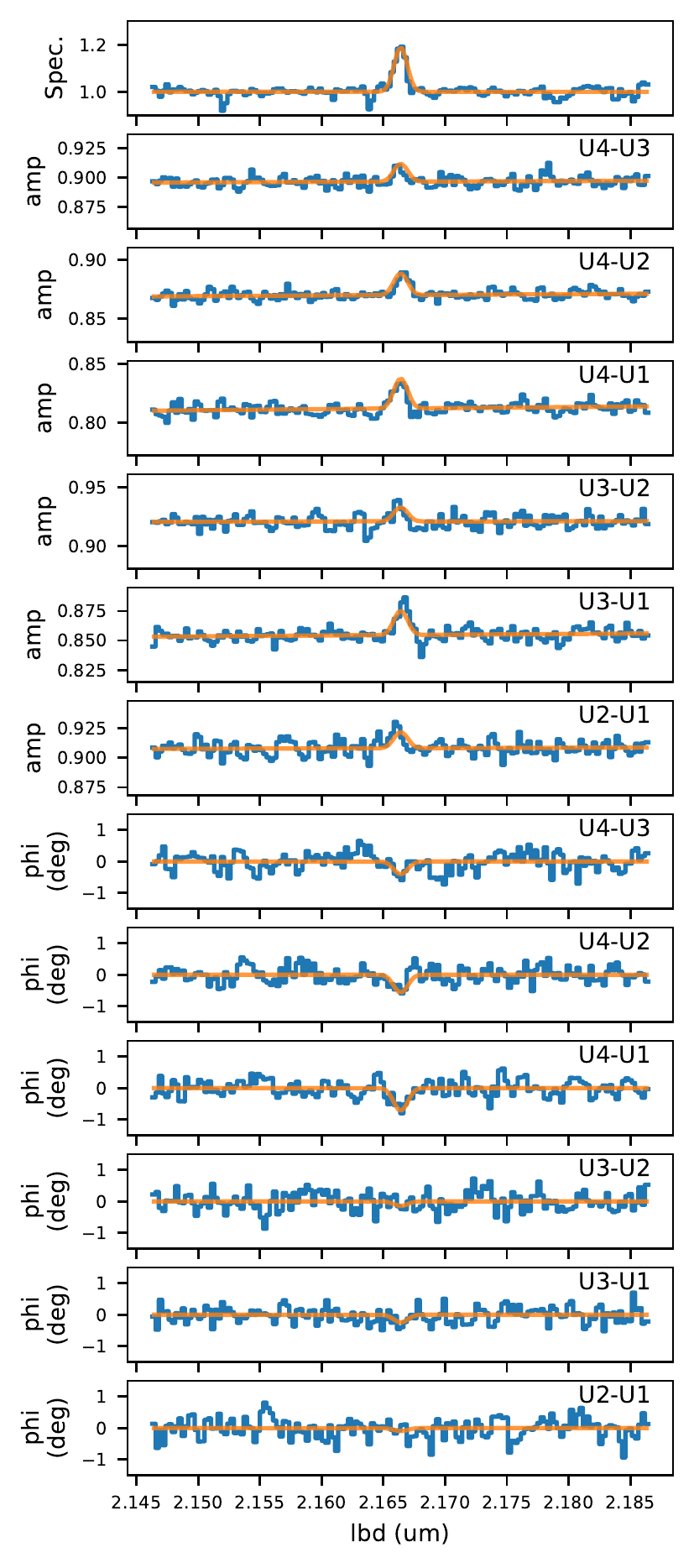}
  \includegraphics[width=0.48\hsize]{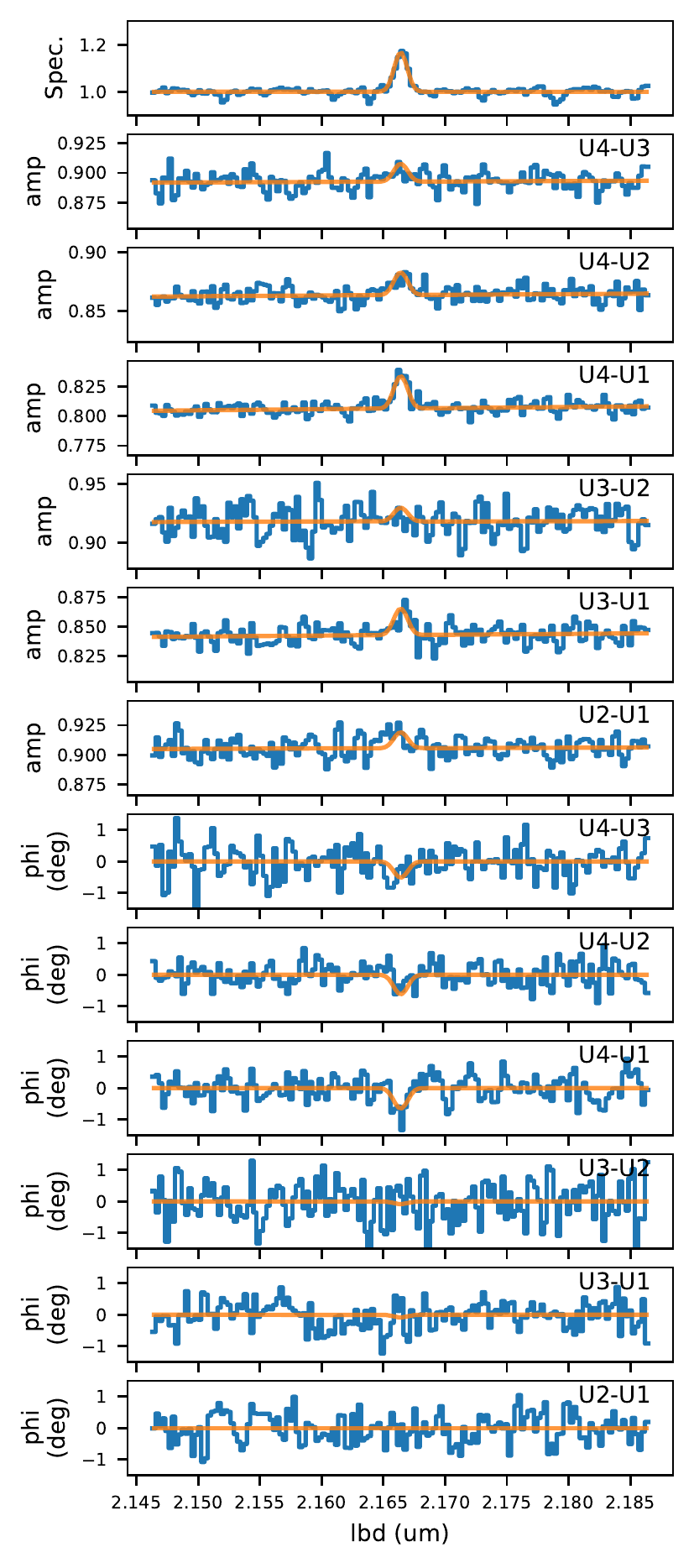}
  \caption{Spectrally dispersed interferometric observables around the Br$\gamma$ line for the night 2019-06-22 (left) and 2019-06-23 (right) along the 6 UT baselines. The top row shows the intensity spectrum where the \brg\ line is clearly visible in emission at 2.166~$\mu$m. The next 6 rows show visibility amplitudes, and the lower 6 differential phases. For all observables, the continuum has been normalized to reproduce the predictions of the model of the FT data, and only the parameters related to the Br$\gamma$ emission are adjusted. The best-fit model is shown in orange.}
  \label{fig:BrG}%
\end{figure*}

Figure~\ref{fig:BrG} shows the spectrally dispersed interferometric quantities for each baseline. 

\section{Models with azimuthal modulation}

By construction, the reference of the differential phases measured across the \brg\ emission line is the barycenter of the continuum emission. Given its small amplitude, a possible alternate explanation for the phase shift detected across the \brg\ emission line is that the continuum emission is not exactly centered on the star, while \brg\ emission is. However, any such model would be intrinsically asymmetric, because the star itself does contribute significantly to the continuum. The upper limits on the phase closures measured on the continuum define an upper limit on such an asymmetry.

\citet{Lazareff17} implements asymmetries by means of azimuthal modulations of the circumstellar emission. Another possibility would be to shift the entire circumstellar emission with respect to the star, but this is practically indistinguishable given our limited spatial resolution. We restrict our analysis to the first terms of the modulation as defined in \citet{Lazareff17}, where the coefficient c1 (resp. s1) measures the flux asymmetry along the major (resp. minor) axis of the disk. We obtained upper limits of $c1<0.3$ and $s1<0.08$. Larger values for these parameters generate phase closure incompatible with the observations. Formally, the best asymmetric model is with $c1=0.15$, $s1=0$, which corresponds to a slight brightening along the major axis of the disk, towards north. This detection is, however, not significant, owing to the fact that the phase closures in the continuum are compatible with zero.

We then performed the analysis of the spectrally dispersed observations described in Sect.~\ref{sec:brG_fit}, but accounting for possible asymmetries in the underlying model of the continuum. Results are summarized in Table~\ref{tab:modulated_models}. The offset of the \brg\ line emitting region with respect to the center of the star indeed depends on the amount of asymmetry in the continuum, because the latter shifts the reference of the phase measurements. In summary, we conclude that the offset of the \brg\ line emitting region with respect to the center of the star is affected by an additional $\pm15\,\mu$as uncertainty when accounting for the upper limits we derive on the asymmetry of the continuum emission.

\begin{table}[h]
  \caption{Best-fit offset of the \brg\ line-emitting region with respect to the center of the star when accounting for azimuthal modulations in the model of the continuum.}
  \label{tab:modulated_models}
  \centering
  \vspace{0.1cm}
  \begin{tabular}{l l l}
    \hline \hline
    c1 & s1     & Offset\\
    0    & 0    & 52$\,\mu$as \\
    0.15 & 0    & 43.6$\,\mu$as \\
    0.3  & 0    & 37.8$\,\mu$as \\
    0    & 0.08 & 48.7$\,\mu$as \\
    \hline
  \end{tabular}
\end{table}

\end{document}